\documentclass[conference]{IEEEtran}
\IEEEoverridecommandlockouts

\usepackage{cite}
\usepackage{amsmath,amssymb,amsfonts}
\usepackage{algorithmic}
\usepackage{graphicx}
\usepackage{textcomp}
\usepackage{xcolor}
\usepackage{pdfpages}
\def\BibTeX{{\rm B\kern-.05em{\sc i\kern-.025em b}\kern-.08em
    T\kern-.1667em\lower.7ex\hbox{E}\kern-.125emX}}
\begin{document}

\title{A Near-Field Compatible Model\\ for 2D Waveguide-Fed Metasurfaces\vspace{-0.5cm}}
\author{Panagiotis Gavriilidis and George C. Alexandropoulos
\\
\IEEEauthorblockA{Department of Informatics and Telecommunications, National and Kapodistrian University of Athens, Athens 16122, Greece}
\\
\vspace{-0.4 cm}
\IEEEauthorblockA{emails: \{pangavr, alexandg\}@di.uoa.gr}
\vspace{-1 cm}
\thanks{
P. Gavriilidis acknowledges the support of the Fulbright Program and the Fulbright Foundation in Greece through a Fulbright Fellowship.}
}

\maketitle
\begin{abstract}
This paper presents a novel physically consistent analytical model for two-dimensional (2D) waveguide-fed metasurface antennas that is based on the discrete dipole approximation. The proposed framework extends previous works deriving power-consistent constraints on the magnetic polarizability tensor, leading to closed-form expressions for the effective polarizabilities. The model is validated through full-wave simulations for multi-feed settings, and is extended to a near-field compatible~formulation enabling accurate predictions in the radiating near-field.

\end{abstract}

\begin{IEEEkeywords}
Dynamic metasurface antennas, discrete dipole approximation, 2D waveguide, mutual coupling, polarizability.
\end{IEEEkeywords}
\vspace{-0.2 cm}
\section{Introduction}
To meet the high-throughput demands of future wireless networks, new antenna
architectures beyond fully digital Multiple-Input Multiple-Output (MIMO)
arrays are being actively investigated.
In particular, future systems are expected to transition toward hybrid
analog-digital solutions to reduce hardware complexity and power
consumption.
A further step in this direction is the use of metasurface-based arrays,
which replace conventional RF circuitry, such as phase shifters, with
reconfigurable metamaterial elements, enabling improved energy efficiency
and large-scale deployment \cite{Basar2024emergingHW}.
Dynamic Metasurface Antennas (DMAs), composed of tunable metamaterials at the
RF front-end and driven by a limited number of digital feeds, constitute a
promising realization of this paradigm \cite{Shlezinger2021Dynamic}.
Unlike conventional phased arrays, where phase shifters and power splitters
are employed at the circuit level, DMAs replace these components with
densely packed metamaterial elements excited through waveguide-based
feeding mechanisms, which necessitates accurate antenna-level
electromagnetic modeling.
As a result, mutual coupling effects are significantly more pronounced, due
both to the lack of isolation structures and to the strong interaction
between the elements through guided and radiated fields \cite{pulidomancera2018,williams2022electromagnetic,gavras20262D_DMA}.
Conventional DMA implementations typically rely on multiple stacked One-Dimensional (1D) waveguides or microstrip lines, each fed through a dedicated excitation port, with metamaterial elements etched on top \cite{Shlezinger2021Dynamic}. In such architectures, scalability remains limited by the number of 1D waveguides and associated feeds. In contrast, the adoption of a 2D waveguide or else a Parallel Plate Waveguide (PPW), allows the entire aperture to be excited using a small number of feeds, potentially even a single one, leading to highly energy-efficient and naturally scalable designs by simply increasing the plate size \cite{pulidomancera2018}. However, for PPW-based metasurfaces, simple transmission-line models
(e.g.,~\cite{davidsmith2017}) are no longer applicable, as one must account for radial guided-wave propagation and strong mutual coupling between elements through the PPW and free space. These challenges were addressed in \cite{pulidomancera2018} through a discrete-dipole–based analytical model for PPW-fed metasurfaces.
In this paper, in contrast to \cite{pulidomancera2018}, we derive power-conservation constraints on the element polarizabilities, leading to
closed-form effective polarizability expressions without relying on numerical extraction procedures.
Moreover, we extend the model to support multiple excitation feeds and a Near-Field (NF) compatible formulation, and validate it against Full-Wave (FW)
simulations in both the NF and Far-Field (FF) regimes.

\vspace{-0.2 cm}

\section{Modeling of 2D Waveguide-Fed Metasurfaces}
\vspace{-0.2 cm}
We consider a Transmitter (TX) employing a metasurface-based antenna, realized through a PPW-fed structure with plate separation \(h\), filled with air (\cite[Fig.~2]{pulidomancera2018}). To accurately model this architecture, we adopt the discrete dipole approximation, where the subwavelength size of each metamaterial element justifies its representation as a magnetic dipole \cite{williams2022electromagnetic,pulidomancera2018,davidsmith2017}. Each element is described by a polarizability tensor that links the induced three-dimensional magnetic field to the dipole moment vector. In the present case, we restrict our attention to elements with a \(2 \times 2\) sized polarizability tensor, as the third dimension cannot be excited because the magnetic field perpendicular to the plate is zero due to the boundary conditions on the waveguide walls. Without loss of generality, we consider the metasurface to lie in the $x$–$y$ plane at $z=0$, thus the polarizability tensor for each $n$-th element ($n=1,2,\ldots,N$, with $N$ denoting the total number of elements) is characterized by a polarizability matrix: \(\mathbf{A}_n=[\alpha_n^{xx} , \alpha_n^{xy};\alpha_n^{yx} , \alpha_n^{yy}]\in \mathbb{C}^{2\times 2}\),
where \(\alpha^{xx}_n\) and \(\alpha^{yy}_n\) are the self-polarization terms, while \(\alpha^{xy}_n\) and \(\alpha_n^{yx}\) are the cross-polarization terms. 
In practice, \(\mathbf{A}\) is frequency dependent, i.e., \(\mathbf{A}_n=\mathbf{A}_n(f)\), but, for notation simplicity, we do not explicitly indicate \(f\). 

Denoting by $\mathbf{m}_n=[m_n^x,m_n^y] \in \mathbb{C}^{2 \times 1}$ the magnetic dipole moment of the $n$-th element and by $\mathbf{h}_{\text{loc},n}=[h^x_{\text{loc},n},h^y_{\text{loc},n}]\in \mathbb{C}^{2 \times 1}$ the local magnetic field at its position, the constitutive relation is given by:
\vspace{-0.3 cm}
\begin{equation}\label{eq: magnetic moment from local field}
    \mathbf{m}_n = \mathbf{A}_n \mathbf{h}_{\text{loc},n}.
\end{equation}
The local field acting on the $n$-th element is given by the sum of the guided excitation $\mathbf{h}_{0,n}=[h^{x}_{0,n},h^y_{0,n}]\in \mathbb{C}^{2 \times 1}$ from the feed and the scattered fields generated by all other dipoles:
\vspace{-0.2 cm}
\begin{equation}
    \mathbf{h}_{\text{loc},n} = \mathbf{h}_{0,n} + 
    \sum_{j=1, j\neq n}^{N}\mathbf{G}_{n,j}\, \mathbf{m}_j,
    \label{eq:Hloc_expanded_final}
\end{equation}
\vspace{-0.1 cm}
where $\mathbf{G}_{n,j}\in \mathbb{C}^{2 \times 2}$ models the interaction between elements $j$ and $n$, while the self-term is excluded. Defining $\mathbf{r}_n \triangleq [r_{x_n},r_{y_n},0]$ as the position vector of the $n$-th element, the interactions are modeled as follows:
\begin{equation}\label{eq: Total Green Function}
\mathbf{G}_{n,j} = 
\begin{cases}
\mathbf{G}_{\text{WG}}\left(\mathbf{r}_n-\mathbf{r}_j\right)+ \mathbf{G}_{\text{FS}}\left(\mathbf{r}_n-\mathbf{r}_j\right), & n\neq j,\\
0, & n=j,
\end{cases}
\end{equation}
with $\mathbf{G}_{\text{WG}}$ describing the coupling through the waveguide and $\mathbf{G}_{\text{FS}}$ the coupling through free space for \(x\) and \(y\) polarized fields. For the considered case of a PPW of height $h$ filled with air, the waveguide contribution is given by\cite{Gavriilidis2026WaveguideMetasurface}:
\vspace{-0.2 cm}
\begin{align}\label{eq: Waveguide Green}
    {G}^{xx}_{\text{WG}}\left(\mathbf{r}_n \text{-}\mathbf{r}_j\right)
    \!= & \frac{\text{-}\jmath k^2}{8h}\!\left[
        H^{(2)}_{0}\!\left(k\rho_{n,j}\right)\!-\! \cos\left(2\psi_{n,j}\right)H^{(2)}_{2}\!\left(k\rho_{n,j}\right)\right]\!,\nonumber\\
    {G}^{xy}_{\text{WG}}\left(\mathbf{r}_n\text{-}\mathbf{r}_j\right)\!
    = & \frac{\text{-}\jmath k^2}{8h} \sin\!\left(2\psi_{n,j}\right)\,
        H^{(2)}_{2}\!\left(k\rho_{n,j}\right),\\
    {G}^{yy}_{\text{WG}}(\cdot)
    \!=\! & \,{G}^{xx}_{\text{WG}}(\cdot)\,\text{and}\;{G}^{xy}_{\text{WG}}(\cdot)
    \!=\! {G}^{yx}_{\text{WG}}(\cdot).\nonumber
\end{align}
The dyadic Green's function that relates the magnetic moment to magnetic field in free-space is given via \cite[eq.~(8.55)]{Novotny_Hecht_2006}:
\begin{align}
    & \mathbf{G}_{\text{FS}}(\mathbf{r}_n\text{-}\mathbf{r}_j) = \bigg[
    \left(\frac{3}{k^2 \rho_{n,j}^2}+\frac{3\jmath}{k\rho_{n,j}}-1\right)\mathbf{R}\label{eq: Free Space Green}\\
    & \quad \qquad \qquad\qquad+\left(1-\frac{\jmath}{k\rho_{n,j}}-\frac{1}{k^2 \rho_{n,j}^2}\right)\mathbf{I}_2
    \bigg]\frac{k^2e^{-\jmath k\rho_{n,j}}}{2\pi\rho_{n,j}},\nonumber\\
     &\mathbf{R}\triangleq \begin{bmatrix}
\cos^2(\psi_{n,j}) & \cos(\psi_{n,j})\sin(\psi_{n,j}) \\
\cos(\psi_{n,j})\sin(\psi_{n,j}) & \sin^2(\psi_{n,j})
\end{bmatrix},
\end{align}
where $H^{(2)}_{\nu}(\cdot)$ denotes the Hankel function of the second kind and order $\nu$\cite[eqs.~(V-14) and (V-15)]{balanis2016antenna}, $k$ is the propagation constant of the guided mode, $\rho_{n,j}\triangleq\lvert\mathbf{r}_n-\mathbf{r}_j\rvert=\sqrt{(r_{x_n}-r_{x_j})^2+(r_{y_n}-r_{y_j})^2}$, \(\mathbf{R}\) is the rotation matrix, and $\psi_{n,j}=\mathrm{atan}\left((r_{y_n}-r_{y_j})/(r_{x_n}-r_{x_j})\right)$. Furthermore, we have accounted for the image of the dipole, created due to the metallic plate that it is placed on, by doubling \(\mathbf{G}_{\rm FS}(\cdot)\).

Then, we extend the formulation of \cite{pulidomancera2018} to the case of multiple excitation feeds. Assuming the feeds are independently controlled ideal thin-wire current sources, the magnetic field induced in the \(n\)-th element due to the \(N_b\) feeds is the superposition of the fields radiated by each feed:  
\vspace{-0.2 cm}
\begin{equation}\label{eq: Excitation Field}
\begin{split}
    h^{x}_{0,n}\!\!=& \frac{\jmath k}{4}\!\sum_{i=1}^{N_b}\!I_i \, H^{(2)}_{1}\!\!\left(k|\mathbf{r}_n-\mathbf{p}_i|\right)\!\sin\!\!\left(\!\!\mathrm{atan}\!\left( \frac{p_{y_i}- r_{y_n}}{p_{x_i}- r_{x_n}}\right)\!\!\right),\\
    h^{y}_{0,n}\!\!=& \frac{\text{-}\jmath k}{4}\!\sum_{i=1}^{N_b}\!I_i \, H^{(2)}_{1}\!\!\left(k|\mathbf{r}_n-\mathbf{p}_i|\right)\!\cos\!\!\left(\!\!\mathrm{atan}\!\left(\! \frac{p_{y_i}- r_{y_n}}{p_{x_i}- r_{x_n}}\!\right)\!\!\right)\!.
\end{split}
\end{equation}
In \eqref{eq: Excitation Field}, \(\mathbf{p}_i\triangleq [p_{x_i},p_{y_i},0]\) denotes the position of the \(i\)-th feed and \(I_i\) represents the externally applied current at the \(i\)-th feed. 

For compactness, the concatenated dipole moment vector is defined as $\mathbf{m} \triangleq [\mathbf{m}_1,\ldots,\mathbf{m}_N]  \in \mathbb{C}^{2N\times 1}$. Analogously, the polarizability matrix containing the polarizabilities of all elements is defined as $\bar{\mathbf{A}} \triangleq {\rm diag} [\mathbf{A}_1,\ldots,\mathbf{A}_N] \in \mathbb{C}^{2N\times 2N}$, i.e., \([\bar{\mathbf{A}}]_{2n-1:2n,2n-1:2n}=\mathbf{A}_n\), and similarly we define the interaction array \(\bar{\mathbf{G}}\in \mathbb{C}^{2 N \times 2 N}\) for which \([\bar{\mathbf{G}}]_{2n-1:2n,2j-1:2j}=\mathbf{G}_{n,j}\), \(\forall \,n,j=1,\ldots,N\). Additionally, the excitation-field vector is expressed as $\mathbf{h}_0 \triangleq [\mathbf{h}_{0,1},\ldots,\mathbf{h}_{0,N}] \in \mathbb{C}^{2N\times 1}$. With these definitions, \eqref{eq: magnetic moment from local field} and \eqref{eq:Hloc_expanded_final} hold for all elements simultaneously, and by solving collectively for the stacked dipole moment vector, yields: \(
    \mathbf{m} = \left(\bar{\mathbf{A}}^{-1} - \bar{\mathbf{G}}\right)^{-1}\mathbf{h}_0
\).

Thereon, we can also distinguish the contribution of the source currents \(I_i\), by introducing the matrix \(\mathbf{H}_f\in\mathbb{C}^{2N \times N_b}\) with \([\mathbf{H}_f]_{2n-1,i}=\frac{\jmath k}{4}H^{(2)}_{1}\!\left(k|\mathbf{r}_n-\mathbf{p}_i|\right)\sin\!\left(\mathrm{atan}\!\left( \frac{p_{y_i}- r_{y_n}}{p_{x_i}- r_{x_n}}\right)\right)\) and \([\mathbf{H}_f]_{2n,i}=\frac{-\jmath k}{4}H^{(2)}_{1}\!\left(k|\mathbf{r}_n-\mathbf{p}_i|\right)\cos\!\left(\mathrm{atan}\!\left( \frac{p_{y_i}- r_{y_n}}{p_{x_i}- r_{x_n}}\right)\right)\). Finally, we define the currents' vector \(\mathbf{i}\triangleq [I_1,\ldots,I_{N_b}]\in \mathbb{C}^{N_b\times 1}\), yielding:
 \(   \mathbf{h}_0 = \mathbf{H}_f \mathbf{i}\).
Combining the latter with \(\mathbf{m}\)'s solution, we can provide a compact characterization of the coupled-dipole physics for 2D waveguide-fed metasurfaces. The final expression linking dipole moments to electric currents in the sources is:
\begin{equation}\label{eq: Dipole Moment from currents}
    \mathbf{m} = \left(\bar{\mathbf{A}}^{-1} - \bar{\mathbf{G}}\right)^{-1}\mathbf{H}_f \mathbf{i}.
\end{equation}

\subsection{Power Conservation Constraint}\label{subsec: Power Conservation Magnetic Dipoles}
Since no amplifying elements are considered, the magnetic polarizability tensor of each dipole must satisfy a passivity constraint following the conservation of energy principle: the power supplied to the $n$-th dipole, $P_{\mathrm{sup},n}$, cannot be smaller than its radiated power, $P_{\mathrm{rad},n}$. In our formulations, we assume time-harmonic fields with the convention $\exp(\jmath\omega t)$. For the latter convention, the Lorentz-oscillator model implies that the imaginary part of the polarizability is negative. 
The physics behind the latter condition will become apparent in the sequel, where it becomes a necessary condition so that the supplied power in the dipole is positive. When matrices are in place instead of scalar polarizability values, this generalizes to the imaginary part of the polarizability matrix, \(\rm Im\{\mathbf{A}_n\}\), being a negative definite matrix. 

From the Poynting theorem \cite[eq.~(4.79)]{tretyakov2003analytical}, the power absorbed by the dipole is derived by integrating over the effective aperture $A$ of the dipole:
\vspace{-0.2 cm}
\begin{equation}\label{eq:Poynting_Thm}
P_{\mathrm{sup},n}
= \tfrac{1}{2}{\rm Re}\!\left\{\int_{A} \mathbf{j}^{\rm T}_n(\mathbf r)\,\mathbf{h}_{\mathrm{loc},n}^{*}(\mathbf r)\,d\mathbf r\right\},
\end{equation}
where $\mathbf{j}_n(\mathbf r)=\jmath\omega\mu_0 (m_n^x\delta(x-r_{x_n})\hat{\mathbf{x}} +m_n^y\delta(y-r_{y_n})\hat{\mathbf{y}} )$ is the magnetic current density \cite[Eq.~(8.49)]{Novotny_Hecht_2006}, with \(\hat{\mathbf{x}}\triangleq[1 ,0]\) and \(\hat{\mathbf{y}}\triangleq[0,1]\) being the unitary vectors of the \(x\) and \(y\) axes.  
Due to the delta distribution the integration collapses to computing a single term in the dipoles position, hence, \eqref{eq:Poynting_Thm} yields:
\begin{equation}\label{eq:Psup_final}
P_{\mathrm{sup},n}
= -\frac{\omega\mu_0}{2}\,\mathbf{h}^{\rm H}_{\text{loc},n}{\rm Im}\left\{\mathbf{A}_n\right\}\mathbf{h}_{\text{loc},n}.
\end{equation}
 Consequently, for the supplied power to be positive, the imaginary part of \(\mathbf{A}_n\)  needs to be a negative definite matrix:  ${\rm Im}\{\mathbf{A}_n\}<\mathbf{0}$. On the other hand, to compute the radiated power, the Poynting theorem is used again, but the induced magnetic field is substituted by the scattered field of the $n$-th element, and the sign is changed, since now the power flowing out of the aperture is computed: \(P_{{\rm rad},n}=\text{-}0.5{\rm Re}\!\left\{\int_{V} \mathbf{j}^{\rm T}_n(\mathbf r)\,\mathbf{h}_{\mathrm{sc},n}^{*}(\mathbf r)\,d\mathbf r\right\}\). The scattered field is given as $\mathbf{h}_{\mathrm{sc},n}(\mathbf r)= \mathbf{G}(\mathbf r\text{-}\mathbf r_n)\mathbf{m}_n$, with $\mathbf{G}(\cdot)=\mathbf{G}_{\mathrm{FS}}(\cdot)+\mathbf{G}_{\mathrm{WG}}(\cdot)$, which yields:
\begin{equation}\label{eq:Prad_1}
P_{\mathrm{rad},n}
= -\tfrac{\omega\mu_0}{2}
\mathbf{h}^{\rm H}_{\text{loc},n}\mathbf{A}_n^{\rm H}{\rm Im}\!\left\{\mathbf{G}(0)\right\}\mathbf{A}_n\mathbf{h}_{\text{loc},n}.
\end{equation}
\(P_{{\rm rad},n}\) is positive since \({\rm Im}\{\mathbf{G}(0)\}\) is a negative definite matrix. 
Evaluating the singularity terms in the four entries of the free space Green's function yields ${\rm Im}\{\mathbf{G}_{\mathrm{FS}}(0)\}=-k^3/(3\pi)\mathbf{I}_2$ \cite{williams2022electromagnetic}. For the waveguide term, the singular parts of the Hankel functions are computed, yielding ${\rm Im}\{\mathbf{G}_{\mathrm{WG}}(0)\}=-k^2/(8h)\mathbf{I}_2$, where $h$ is the plate separation \cite{gavras20262D_DMA}. Hence, combining the above derivations the imaginary part of the self-term of the 2D Green's matrix is given as:
\(    {\rm Im}\{\mathbf{G}(0)\} = -\left(k^3/(3\pi)+k^2/(8h)\right)\mathbf{I}_2\).

Imposing energy conservation for a passive element,
\(
P_{\mathrm{sup},n} \geq P_{\mathrm{rad},n} ,
\)
and substituting \eqref{eq:Psup_final} and \eqref{eq:Prad_1} yields:
\[
\mathbf{A}_n^{\rm H}{\rm Im}\left\{\mathbf{G}(0)\right\}\mathbf{A}_n - {\rm Im}\left\{\mathbf{A}_n\right\} \geq \mathbf{0}
\]
Using the matrix identity ${\rm Im}\{\mathbf{A}_n\}=-\mathbf{A}_n{\rm Im}\{\mathbf{A}^{-1}_n\}\mathbf{A}_n^{\rm H}$, the passivity constraint can be rewritten as:
\begin{equation}\label{eq: passivity_constraint}
{\rm Im}\{ \mathbf{A}_n^{-1} \}+{\rm Im}\{\mathbf{G}(0)\}\geq \mathbf{0}.
\end{equation}
Condition \eqref{eq: passivity_constraint} is interpreted as: the imaginary part of \(\mathbf{A}^{-1}+\mathbf{G}(0)\) must be a semi positive definite matrix. This condition is general for other architectures, not restricted to the 2D waveguide presented here, but the exact values of \(\mathbf{G}(0)\) are computed for a specific setup, e.g., the PPW for this case.

Finally, instead of enforcing \eqref{eq: passivity_constraint} as an explicit inequality, one can build it directly into the definition of the polarizability via the standard Radiation--Reaction (RR) correction. We distinguish between an intrinsic polarizability $\mathbf{A}^{'}_n$, which is obtained from a quasi-static model (such as \cite[eq.~(27)]{mancera2017polarizability}) and does \emph{not} include the dipole’s self-field, and the effective polarizability $\mathbf{A}_n$, which does. Essentially, the intrinsic polarizability is the property of the element due to its geometry without accounting for the radiation environment, while the effective polarizability accounts for the environment (the PPW in our case). For a dipole embedded in the PPW, the self-interaction is represented by the Green’s function evaluated at the dipole location, so the RR corrected polarizability is written as: 
\vspace{-0.2 cm}
\begin{equation}\label{eq: rr_correction}
\mathbf{A}_n
= \mathbf{A}^{'}_n \left(\mathbf{I}_2 - \jmath {\rm Im}\{\mathbf{G}(0)\}\mathbf{A}^{'}_n\right)^{-1},
\end{equation}
which is the matrix equivalent of \cite[eq.~(1)]{mancera2017polarizability}.
Taking imaginary parts in \eqref{eq: rr_correction}, and  after performing some mathematical manipulations, yields:  
\(
{\rm Im}\{\mathbf{A}^{-1}_n\}={\rm Im}\{(\mathbf{A}^{'})^{-1}\} - {\rm Im}\{\mathbf{G}(0)\}
\).
Since for a passive element it holds ${\rm Im}\{(\mathbf{A}^{'})^{-1}\}> \mathbf{0}$; then it directly applies that:  
\(
{\rm Im}\{\mathbf{A}_n^{-1}\} + {\rm Im}\{\mathbf{G}(0)\}
\;\geq\;
\mathbf{0},
\)
which is exactly the passivity condition \eqref{eq: passivity_constraint}. In the lossless limit, ${\rm Im}\{\mathbf{A}_n^{'}\}=0$ the corrected polarizability \eqref{eq: rr_correction} saturates the bound, i.e., all dissipation is purely radiative.

\begin{figure*}[t]
    \includegraphics[width=1\linewidth]{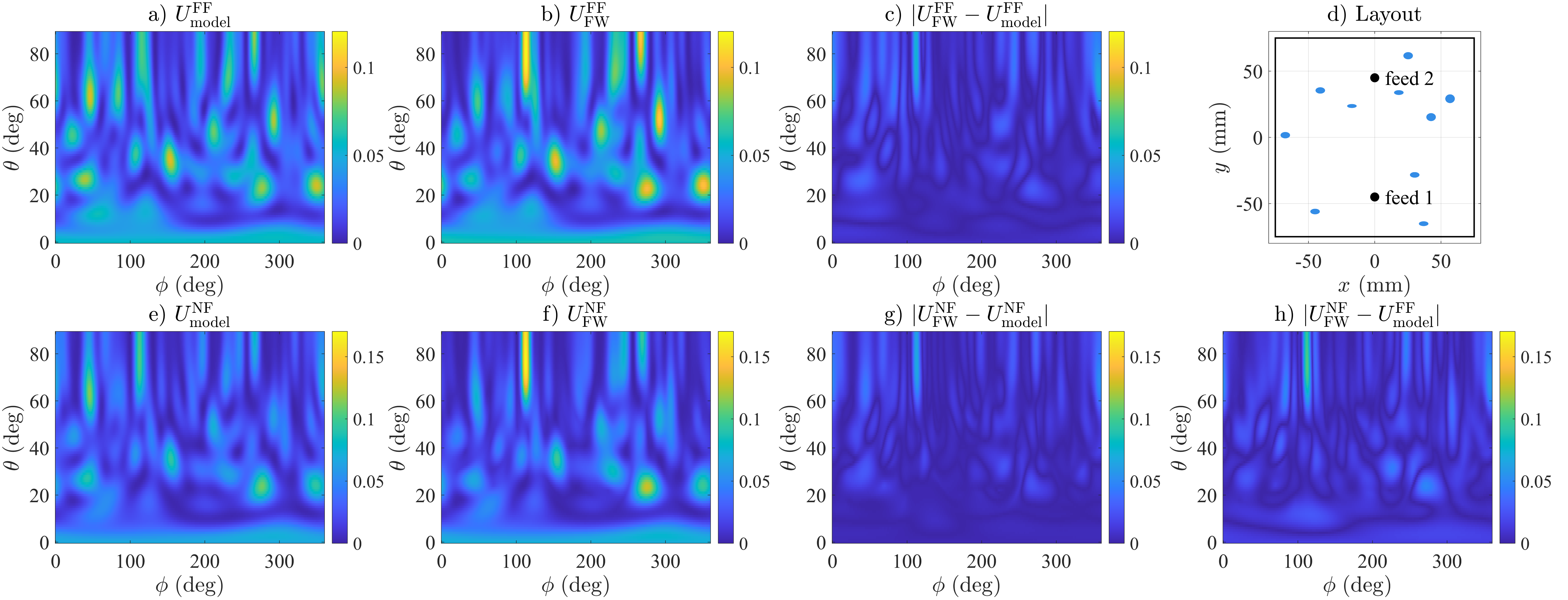}
    \vspace{-0.7 cm}
    \caption{Radiation intensity comparisons: (a)–(c) show the FF intensities obtained from the analytic model, FW simulations, and their difference, respectively; (e)–(g) show the corresponding NF quantities at $R_{\ell}=0.4$~m. Plot (h) compares the NF FW result $U_{\rm FW}^{\rm NF}$ with the FF model prediction $U_{\rm model}^{\rm FF}$. Plot (d) shows the top view of the PPW used in the simulations, where blue ellipses denote the elements and black circles the feeds.
}
    \vspace{-0.4 cm}
    \label{fig: all figures}
\end{figure*}

\subsection{Integration in MIMO Models}
\subsubsection{MIMO Channel}
By solving the radiation problem in the domain outside of the waveguide the electric field \(\mathbf{e}_{{\rm sc},n} \in \mathbb{C}^{2 \times 1}\) radiated from each \(n\)-th metamaterial can be acquired as a function of its magnetic moment \(\mathbf{m}_n\). The authors in \cite{pulidomancera2018} derived the electric field in free-space using the FF approximation. In this work, we do not make the complete FF approximation, but only eliminate terms divided by the square distance \(R^2_n\) between the \(n\)-th element and the observation point as per the radiative NF approximation. We compute the electric field in free space  \(\mathbf{e}_{{\rm sc},n}(\mathbf{r})=\jmath \omega \mu_0 [\nabla \!\times \!\tilde{\mathbf{G}}_{\rm FS}(\mathbf{r}-\mathbf{r}_n)]\mathbf{m}_n\), where \(\tilde{\mathbf{G}}_{\rm FS}\in \mathbb{C}^{3 \times 3}\) is the FF dyadic Green function as given in \cite[eq.~(8.61)]{Novotny_Hecht_2006}, but doubled due to the dipole image created by the top plate of the waveguide \footnote{To circumvent the size incompatibility issue in the \([\nabla\!\times\! \tilde{\mathbf{G}}(\mathbf{r}\text{-}\mathbf{r}_n)]\mathbf{m}_n\) multiplication, a zero needs to be padded to \(\mathbf{m}_n\).}. Then, the electric field is transformed from Cartesian to spherical coordinates, since the use of spherical coordinates eases the analysis, as the longitudinal component of both the electric and magnetic fields (i.e., the component parallel to propagation) is zero. Hence, only the azimuth, \(\phi\), and elevation, \(\theta\), components are of interest. Consequently, the electric field at an observation point \(\mathbf{s}_{\ell}\) due to the dipole located at \(\mathbf{r}_n\), \(\mathbf{e}_{{\rm sc},n}(\mathbf{s}_{\ell}-\mathbf{r}_n)\), which for ease of notation we write just as \(\mathbf{e}_{{\rm sc},n}(\mathbf{s}_{\ell})\), has the following two components:
\begin{align}\label{eq: electric_field_due_to_nth_dipole}
\!\!e_{{\rm sc},n}^{\theta}(\mathbf{s}_{\ell})\!
&=\!
\frac{\eta k^{2}e^{-\jmath kR_{n,\ell}}}{2\pi  R_{n,\ell}}\!
\big(\!m_{n}^{x}\sin(\!\phi_{n,\ell}\!) \!-\! m_{n}^{y}\cos(\!\phi_{n,\ell}\!)\!\big),\\[2mm]
\!\! e_{{\rm sc},n}^{\phi}(\mathbf{s}_{\ell})
\!&=\!
\frac{\eta k^{2}e^{-\jmath kR_{n,\ell}}}{2\pi R_{n,\ell}}
\!\big(\!m_{n}^{x}\cos(\!\phi_{n,\ell}\!) \!+ \!m_{n}^{y}\sin(\!\phi_{n,\ell}\!)\!\big)\!\cos(\!\theta_{n,\ell}\!),\nonumber
\end{align}
where \(\eta\) is the free space impedance, \(R_{n.\ell}\) is the radial distance between \(\mathbf{s}_{\ell}\) and \(\mathbf{r}_n\), and \(\theta_{n,\ell} ={\rm acos}\left(({s_{z_{\ell}} - r_{{z}_n}})/{R_{n,\ell}}\right)\) as well as \(\phi_{n,\ell} = {\rm atan}\left( ({s_{y_{\ell}}- r_{y_n}})/({s_{x_{\ell}}- r_{x_n}})\right)\) are the elevation and azimuth angles between them. 
In analogy with the focusing vector in NF MIMO systems, the electric field components scattered toward the observation point $\mathbf{s}_{\ell}$ can be written in a linear form with respect to the stacked
magnetic moments $\mathbf{m}\in\mathbb{C}^{2N\times 1}$. However, due to the NF formulation, the elevation and azimuth angles between each \(n\)-th dipole and the observation point, $\theta_{n,\ell}$ and $\phi_{n,\ell}$, depend on the specific dipole location
$\mathbf{r}_n$, and thus the field components $e_{{\rm sc},n}^{\theta}(\mathbf{s}_{\ell})$ and $e_{{\rm sc},n}^{\phi}(\mathbf{s}_{\ell})$ in \eqref{eq: electric_field_due_to_nth_dipole} are expressed in
dipole-dependent local spherical bases. As a result, these terms do not correspond to a common polarization basis and cannot be directly summed.

To enable coherent superposition, each dipole contribution is projected onto a common
spherical basis associated with the observation point $\mathbf{s}_\ell$ and defined with
respect to the TX center. Let $(\theta_\ell,\phi_\ell)$ denote the elevation and azimuth
angles of $\mathbf{s}_\ell$ with respect to the TX center, and let
$\hat{\boldsymbol{\theta}}_\ell$ and $\hat{\boldsymbol{\phi}}_\ell$
denote the corresponding spherical-to-Cartesian projection vectors,
defined as
$\hat{\boldsymbol{\theta}}_\ell
\triangleq
[\cos(\theta_\ell)\cos(\phi_\ell),\,
 \cos(\theta_\ell)\sin(\phi_\ell),\,
 -\sin(\theta_\ell)]^{\mathrm T}$
and
$\hat{\boldsymbol{\phi}}_\ell
\triangleq
[-\sin(\phi_\ell),\,
 \cos(\phi_\ell),\,
 0]^{\mathrm T}$.
Similarly, $\hat{\boldsymbol{\theta}}_{n,\ell}$ and
$\hat{\boldsymbol{\phi}}_{n,\ell}$ denote the spherical-to-Cartesian
projection vectors associated with the direction
$\mathbf{s}_\ell-\mathbf{r}_n$, obtained by replacing
$(\theta_\ell,\phi_\ell)$ with $(\theta_{n,\ell},\phi_{n,\ell})$ in the
above definitions. Then, we can define the $2\times2$ projection matrix:
\vspace{-0.1 cm}
\begin{equation}\label{eq:Tnell_def}
\mathbf T_{n\rightarrow \ell}
\triangleq
\begin{bmatrix}
\hat{\boldsymbol{\theta}}_\ell^{\mathrm T}\hat{\boldsymbol{\theta}}_{n,\ell} &
\hat{\boldsymbol{\theta}}_\ell^{\mathrm T}\hat{\boldsymbol{\phi}}_{n,\ell}\\
\hat{\boldsymbol{\phi}}_\ell^{\mathrm T}\hat{\boldsymbol{\theta}}_{n,\ell} &
\hat{\boldsymbol{\phi}}_\ell^{\mathrm T}\hat{\boldsymbol{\phi}}_{n,\ell}
\end{bmatrix},
\end{equation}
which maps the transverse components of the \(n\)-th dipole, \(e_{{\rm sc},n}^{\phi}(\mathbf{s}_{\ell})\) and \(e_{{\rm sc},n}^{\theta}(\mathbf{s}_{\ell})\), from their local basis into the common basis, by projecting through $\mathbf T_{n\rightarrow \ell}\,\forall n$. To express all dipole contributions in the common spherical basis, we first introduce the local focusing vectors $\mathbf a^{\theta}(\mathbf s_\ell)$ and
$\mathbf a^{\phi}(\mathbf s_\ell)$, which collect the dipole-dependent responses prior to basis unification and are defined element-wise
for $n=1,\ldots,N$ as:
\vspace{-0.2 cm}
\begin{align}
\begin{alignedat}{2}\label{eq:focusing_vector_definitions}
    &\big[\mathbf{a}^{\theta}(\mathbf{s}_{\ell})\big]_{2n-1} &&= 1/R_{n,\ell}\,\sin(\phi_{n,\ell})\,e^{-jkR_{n,\ell}}, \\ 
    &\big[\mathbf{a}^{\theta}(\mathbf{s}_{\ell})\big]_{2n}   &&= -1/R_{n,\ell}\,\cos(\phi_{n,\ell})\,e^{-jkR_{n,\ell}}, \\
    &\big[\mathbf{a}^{\phi}(\mathbf{s}_{\ell})\big]_{2n-1}  &&= 1/R_{n,\ell}\,\cos(\phi_{n,\ell})\,\cos(\theta_{n,\ell})\,e^{-jkR_{n,\ell}}, \\
    &\big[\mathbf{a}^{\phi}(\mathbf{s}_{\ell})\big]_{2n}    &&= 1/R_{n,\ell}\,\sin(\phi_{n,\ell})\,\cos(\theta_{n,\ell})\,e^{-jkR_{n,\ell}}.
\end{alignedat}
\end{align}
Consequently, to account for the basis mismatch, the focusing vectors are transformed
element-wise through $\mathbf T_{n\rightarrow\ell}$, yielding the
projected (common-basis) focusing vectors
$\tilde{\mathbf a}^{\theta}(\mathbf s_\ell)\in\mathbb C^{2N\times 1}$ and
$\tilde{\mathbf a}^{\phi}(\mathbf s_\ell)\in\mathbb C^{2N\times 1}$, defined
for $n=1,\ldots,N$ as:
\begin{equation}\label{eq:projected_focusing_vectors}
\begin{alignedat}{2}
&\begin{bmatrix}
    \big[\tilde{\mathbf a}^{\theta}(\mathbf s_\ell)\big]_{2n-1}\\
 \big[\tilde{\mathbf a}^{\phi}(\mathbf s_\ell)\big]_{2n-1}   
\end{bmatrix}
&&=\mathbf{T}_{n\rightarrow \ell}\begin{bmatrix}
\big[\mathbf a^{\theta}(\mathbf s_\ell)\big]_{2n-1}\\
\big[\mathbf a^{\phi}(\mathbf s_\ell)\big]_{2n-1}
\end{bmatrix},\\
&\begin{bmatrix}
    \big[\tilde{\mathbf a}^{\theta}(\mathbf s_\ell)\big]_{2n}\\
 \big[\tilde{\mathbf a}^{\phi}(\mathbf s_\ell)\big]_{2n}   
\end{bmatrix}
&&=\mathbf{T}_{n\rightarrow \ell}\begin{bmatrix}
\big[\mathbf a^{\theta}(\mathbf s_\ell)\big]_{2n}\\
\big[\mathbf a^{\phi}(\mathbf s_\ell)\big]_{2n}
\end{bmatrix}.
\end{alignedat}
\end{equation}
Then, the scattered field components at $\mathbf{s}_\ell$ are given as:
\begin{equation}\label{eq: total electric field_projected}
\begin{aligned}
e_{\rm sc}^{\theta}(\mathbf{s}_{\ell})
&=
{\eta k^{2}}/({2\pi})\,
\tilde{\mathbf a}^{\theta}(\mathbf s_\ell)^{\mathrm T}\mathbf m,\\
e_{\rm sc}^{\phi}(\mathbf{s}_{\ell})
&=
{\eta k^{2}}/({2\pi })\,
\tilde{\mathbf a}^{\phi}(\mathbf s_\ell)^{\mathrm T}\mathbf m .
\end{aligned}
\end{equation}

Using \eqref{eq: total electric field_projected}, we define the dual-polarized channel
matrix between the TX and \(L\) observation points as
\(\mathbf{H} \in \mathbb{C}^{2L\times 2N}\),
with the rows associated with \(\mathbf{s}_{\ell}\), for \(\ell = 1,\ldots,L\),  given by:
\begin{equation}\label{eq: channel_matrix_definition}
\begin{alignedat}{2}
&[\mathbf H]_{2\ell-1,:}
&&=
{\eta k^{2}}/({2\pi})\,
\tilde{\mathbf a}^{\theta}(\mathbf s_\ell)^{\mathrm T},\\
&[\mathbf H]_{2\ell,:}
&&=
{\eta k^{2}}/({2\pi})\,
\tilde{\mathbf a}^{\phi}(\mathbf s_\ell)^{\mathrm T}.
\end{alignedat}
\end{equation}

\subsubsection{TX Modeling} The electric signal received at the \(L\) observation points is given as: \(\mathbf{y} = \mathbf{H}\mathbf{m} + \mathbf{n}\), where \(\mathbf{n}\) is the additive white Gaussian noise. To reformulate the latter w.r.t. the current vector \(\mathbf{i}\), we substitute \(\mathbf{m}\) from \eqref{eq: Dipole Moment from currents}, yielding: 
\vspace{-0.15 cm}
\begin{equation}\label{eq: Received Signal}
    \mathbf{y} = \mathbf{H} \left(\bar{\mathbf{A}}^{-1} - \bar{\mathbf{G}}\right)^{-1}\mathbf{H}_f \mathbf{i} + \mathbf{n}.
\end{equation}
\vspace{-0.05 cm}
Since $\mathbf{i}$ includes both the digital precoding and the transmitted symbol, it can be written as $\mathbf{i}=\mathbf{Vs}$, with $\mathbf{V}$ being the precoder and $\mathbf{s}$ the  information symbol.


\subsubsection{FF Channel as a Special Case}\label{subsec: Far field magnetic}
In the FF regime, the distance between the $n$-th dipole and the
observation point $\mathbf{s}_\ell$ can be approximated as
$R_{n,\ell}\approx R_\ell-\hat{\mathbf u}_\ell^{\mathrm T}\mathbf r_n$,
where $\hat{\mathbf u}_\ell\triangleq
[\sin\theta_\ell\cos\phi_\ell,\,
 \sin\theta_\ell\sin\phi_\ell,\,
 \cos\theta_\ell]^{\mathrm T}$
denotes the propagation direction associated with $\mathbf{s}_\ell$, and \(R_{\ell}\) is the distance between the TX's center and \(\mathbf{s}_{\ell}\).

Under this approximation, the angles
$\theta_{n,\ell}$ and $\phi_{n,\ell}$ become independent of the dipole
index $n$ and satisfy
$\theta_{n,\ell}\approx\theta_\ell$ and
$\phi_{n,\ell}\approx\phi_\ell\,\forall n$, while path loss changes with \(R_{\ell}\) instead of \(R_{n,\ell}\).
Consequently, all dipole contributions are expressed in the same
spherical basis, and the NF projection matrices
$\mathbf T_{n\rightarrow\ell}$ reduce to identity matrices, i.e.,
$\mathbf T_{n\rightarrow\ell}=\mathbf I_2$, yielding \(\tilde{\mathbf{a}}^{\theta}(\cdot)=\mathbf{a}^{\theta}(\cdot)\) as well as \(\tilde{\mathbf{a}}^{\phi}(\cdot)=\mathbf{a}^{\phi}(\cdot)\). Accordingly, the focusing vectors reduce to their FF forms, by replacing in \eqref{eq:focusing_vector_definitions} all \(\phi_{n,\ell}\) and \(\theta_{n,\ell}\) angles with \(\phi_{\ell}\) and \(\theta_{\ell}\), as well as substituting the radial distance \(R_{n,\ell}\) in the exponent with its approximation \(R_\ell-\sin(\theta_{\ell})\cos(\phi_{\ell})r_{x_n}-\sin(\theta_{\ell})\sin(\phi_{\ell})r_{y_n}\).

The corresponding FF dual-polarized channel matrix
$\mathbf H_{\rm FF}\in\mathbb{C}^{2L\times2N}$ is obtained from \eqref{eq: channel_matrix_definition} by replacing the NF focusing vectors with their FF counterparts and omitting the projection matrices. Hence, the proposed NF MIMO channel model naturally reduces to the steering-vector-based formulation as a special case.
\vspace{-0.45 cm}
\section{Validation}
\vspace{-0.2 cm}

To validate the effective polarizability formula as well as the analytic NF and FF channel models, we consider the case of elliptic irises, for which closed-form expressions of the intrinsic polarizabilities exist. We define an elliptic aperture with semi-major axis $l_1$ aligned along the $x$-direction and semi-minor axis $l_2$ aligned along the $y$-direction. The intrinsic polarizabilities for this case can be found in \cite[Table~12.1]{collin1990field}. However, bridging the definition from a scattering object in free space to an aperture connecting two half-spaces (the waveguide and free space) requires a renormalization. Specifically, the intrinsic polarizability of the aperture corresponds to one-fourth of the value for the elliptic iris in free space as detailed in \cite{collin1990field}. Hence, \(\mathbf{A}^{\prime}_n\) is given in~\cite[Ch.~7, eq.~(70b)]{collin1990field}, where due to the symmetry of the iris the cross-polarization terms are zero. Then, to retrieve the effective polarizability \(\mathbf{A}_n\), we employ the RR correction formula derived in \eqref{eq: rr_correction}. 

We examine elliptic iris elements with a fixed semi-major axis $l_1 = 3.6$~mm and varying semi-minor axis $l_2 \leq l_1$. At the operating frequency of $10$~GHz that we consider in this section, the electrical length of the element's major axis corresponds to $k(2l_1) \approx 1.51$. While the RR theory is rigorously derived for electrically small apertures ($k(2l_1) \ll 1$), it is demonstrated in \cite[Figs.~4.14--4.15]{tretyakov2003analytical} that the analytic reaction correction terms begin to deviate from the actual ones when $k(2l_1)$ exceeds $1$ but the approximation remains reasonable till \(1.5\). Consequently, our selected dimension lies at the upper operational limit of this quasi-static approximation.

The dimensions of the PPW are $150\times150~\mathrm{mm}^2$ and plate separation is $h=5.21$~mm.
Two thin electric line sources, modeling ideal current-carrying wires, are
placed inside the waveguide at the locations
$\mathbf{p}_1=[0,\,45]^\mathsf{T}$~mm and
$\mathbf{p}_2=[0,\,-45]^\mathsf{T}$~mm.
Each source is excited with a unit-amplitude current: \(I_1\) and \(I_2=1\)~A.
On the top plate of the PPW, $N=10$ elliptic irises are placed, with different \(l_2\), at
random positions, as illustrated in Fig.~\ref{fig: all figures}(d).

Let $\mathbf{s}_{\ell}$ denote an observation point in free space.
To quantify the angular distribution of the scattered field, we employ the
following intensity-based metric: \(U(\mathbf{s}_{\ell})
=
\frac{R_{\ell}^{2}}{2\eta}
e^{\rm tot}_{\rm sc}(\mathbf{s}_{\ell})^{2},\)
where $e^{\rm tot}_{\rm sc}(\mathbf{s}_{\ell})$ is the magnitude of the total
scattered electric field.
The quantity \(U(\cdot)\) has \(\rm W\) units by construction; in the FF, where the radiated field is transverse, it
coincides with the standard definition of radiation intensity
$U(\theta_{\ell},\phi_{\ell})$ ($\mathrm{W/sr}$).

In the proposed analytic model, the total scattered field
$e^{\rm tot}_{\rm sc}(\mathbf{s}_{\ell})$ is evaluated either using the field expression in \eqref{eq: total electric field_projected}
(NF model) or its FF approximation. For FW simulations, \(U(\cdot)\)
is computed using the absolute value of the scattered electric field exported
per observation angle.
Specifically, we evaluate $U_{\rm FW}$ both at a finite observation distance
of $R_{\ell}=0.4$~m and under an FF approximation.
For the considered PPW aperture, the conventional Fraunhofer distance
$R_{\rm FF}=2D^{2}/\lambda$ is approximately $3$~m, which confirms that
$R_{\ell}=0.4$~m lies well within the radiating NF region.
Accordingly, NF FW results may include radial field components and reactive effects, which are not analytically modeled but are accounted for in FW simulations. In the following, we denote by $U_{\rm FW}$ the metric obtained from FW
simulations and by $U_{\rm model}$ the corresponding metric predicted by the
analytic model, while we use the exponent \(\rm NF\) and \(\rm FF\) for near or far-field, accordingly.
In Fig.~\ref{fig: all figures}(a)–(c), we report the FF intensity metric obtained from the
analytic model, $U^{\mathrm{FF}}_{\rm model}$, the corresponding full-wave
result, $U^{\mathrm{FF}}_{\rm FW}$, and their absolute difference,
$|U^{\mathrm{FF}}_{\rm FW}-U^{\mathrm{FF}}_{\rm model}|$, respectively.
The two angular patterns exhibit close agreement over the entire domain, despite the fact that the proposed model is fully
analytical and does not rely on numerical procedures to extract the element
polarizabilities, unlike \cite{pulidomancera2018}.

The discrepancy between the two patterns increases for large elevation angles, i.e., as $\theta$ approaches $90^{\circ}$. This behavior can be attributed to the fact that the present model neglects the contribution of a normal (along the $z$-direction) electric dipole moment induced on the metasurface elements, whose inclusion is expected to reduce the error in the high-$\theta$ region and is left for future work.
Additional discrepancies between the model and FW results can be attributed to the use of analytic effective polarizabilities near the upper
validity limit of the RR theory. To quantify the overall mismatch between \(U_{\rm model}^{\rm FF}\) and \(U_{\rm FW}^{\rm FF}\),
we compute a \emph{pattern-normalized} solid-angle error, obtained by first
normalizing each pattern by its solid-angle integral and then integrating
the absolute difference over the observation angles using $d\Omega=\sin\theta\,d\theta\,d\phi$. This metric isolates discrepancies in the angular distribution while being
insensitive to global amplitude offsets, yielding an average per angle pattern mismatch of $16.6\%$.

Figures~\ref{fig: all figures}(e)–(g) report the same metrics as in Fig.~\ref{fig: all figures}(a)–(c), but evaluated in
the radiating NF at an observation distance of $R_{\ell}=0.4$~m.
It can be observed that the error between the analytic model and the FW results remains comparable to the FF case, indicating that, despite not employing the full NF formulation, the proposed
model accurately captures NF variations and can be reliably used for NF predictions. In Fig.~\ref{fig: all figures}(h), we compare the NF FW results with the FF model evaluated at the same observation distance. In this case, the discrepancy increases over almost the entire angular
domain, highlighting the benefit of employing the proposed NF model when operating within the radiating NF region. Finally, we summarize these observations by reporting the solid-angle–averaged pattern-normalized errors between $U^{\rm NF}_{\rm FW}$ and
$U^{\rm NF}_{\rm model}$, and between $U^{\rm NF}_{\rm FW}$ and $U^{\rm FF}_{\rm model}$, which amount to $18.3\%$ and $23\%$, respectively. This corresponds to an approximately $25\%$ higher error when the FF model is used in the NF regime. Overall, these results confirm that the proposed analytic framework provides accurate and computationally efficient predictions of both FF and NF scattering characteristics for PPW-fed metasurfaces.
\vspace{-0.1 cm}

\bibliographystyle{IEEEtran}
\vspace{-0.23 cm}
\bibliography{references}
\vspace{-0.07 cm}
\end{document}